\documentclass[prd,aps,preprint,epsfig,floats]{revtex4}

\usepackage{graphicx}

\begin{document}

\preprint{\vbox{
\hbox{UMD-PP-05-040} }}
\title{\Large\bf Leptogenesis, $\mu-\tau$ Symmetry and $\theta_{13}$}
\author{\bf R.N. Mohapatra, S. Nasri and Haibo Yu }

\affiliation{ Department of Physics, University of Maryland, College
Park, MD 20742, USA}

\date{February, 2005}

\begin{abstract}
We show that in theories where neutrino masses arise from  type I
seesaw formula with three right handed neutrinos and where large
atmospheric mixing angle owes its origin to an approximate
leptonic $\mu-\tau$ interchange symmetry, the primordial lepton asymmetry 
of
the Universe, $\epsilon_l$ can be expressed in a simple form in
terms of low energy neutrino oscillation parameters as $\epsilon_l
= (a \Delta m^2_\odot+ b \Delta m^2_A \theta^2_{13})$, where $a$
and $b$ are parameters characterizing high scale physics and are
each of order $\leq 10^{-2} $ eV$^{-2}$. We also find that for the
case of two right handed neutrinos, $\epsilon_l \propto
\theta^2_{13}$ as a result of which, the observed value of baryon to 
photon ratio implies a lower limit on $\theta_{13}$. For specific
choices of the CP phase $\delta$ we find $\theta_{13}$ is predicted to be 
between $0.10-0.15$.
 \end{abstract}
\maketitle
\section{Introduction}
   There may be a deep connection between the origin of matter in the
Universe and the observed neutrino oscillations. This speculation is
inspired by the idea that the heavy right handed Majorana neutrinos that
are added to the standard model for understanding small neutrino
masses via the seesaw mechanism\cite{seesaw} can also explain the origin
of matter via their decay. The mechanism goes as follows\cite{fy}: CP
violation in the same Yukawa
interaction of the right handed neutrinos, which go into giving nonzero
neutrino masses
after electroweak symmetry breaking, lead to a primordial lepton
asymmetry via the out of equilibrium decay $N_R\rightarrow \ell + H$
(where  $\ell$ are the known leptons and $H$ is the standard model Higgs
field). This asymmetry subsequently gets converted to baryon-anti-baryon
asymmetry observed today via the the electroweak sphaleron
interactions\cite{krs}, above $T\geq v_{wk}$ ($v_{wk}$ being the weak
scale). Since this mechanism involves
no new interactions beyond those needed in the discussion of
neutrino masses, one would expect that better
understanding of neutrino mass physics would clarify one of the deepest
mysteries of cosmology both qualitatively as well as quantitatively.
This question has been the subject of many investigations in
recent years\cite{buch,BPD,di,branco,hambye,anjan,sahu,rev} in the context of
different neutrino mass models and many interesting pieces
of information about issues such as the spectrum of right handed
neutrinos, upper limit on the neutrino masses etc have been
obtained. In a recent paper,
\cite{nasri}, two of the authors showed that if one assumes that
the lepton sector of minimal seesaw models has a leptonic
$\mu-\tau$ interchange symmetry\cite{mutau,moh}, then
 one can under certain plausible assumptions indeed predict
the magnitude of the matter-anti-matter asymmetry in terms of low energy
oscillation parameter, $\Delta m^2_\odot$ and a high scale CP phase. The
choice of $\mu-\tau$ symmetry was dictated by the fact that it is the
simplest symmetry of neutrino mass matrix that explains the maximal
atmospheric mixing as indicated by data. Using
present experimental value for $\Delta m^2_\odot$, one obtains the right
magnitude for the baryon asymmetry of the Universe.

The results of the paper \cite{nasri} were derived in the limit that
$\mu-\tau$ interchange symmetry is exact. If however a nonzero value for
the neutrino mixing angle $\theta_{13}$ is detected in future
experiments, this would imply that this symmetry is only approximate.
Also, since in the standard model $\nu_\mu$ and $\nu_\tau$ are
members of the $SU(2)_L$ doublets $L_\mu\equiv (\nu_\mu, \mu)$ and
$L_{\tau}\equiv (\nu_\tau,\tau)$, any symmetry between $\nu_\mu$ and
$\nu_\tau$ must be a symmetry between $L_\mu$ and $L_\tau$ at the
fundamental Lagrangian level. The observed difference between the muon and
tau masses would therefore also imply that the $\mu-\tau$ symmetry has to
be an approximate symmetry.  In view of this, it is important to examine
to what extent the results of Ref.\cite{nasri} carry over to the case when
 the symmetry is approximate. We find two interesting results under some
very general assumtions:
 (i)  a simple
formula relating the lepton asymmetry and neutrino oscillation observables
for the case of three right handed neutrinos, i.e.
$\epsilon_l= (a \Delta m^2_\odot+ b \Delta
m^2_A \theta^2_{13})$ and (ii) a relation of the form
$\epsilon_l\propto \theta^2_{13}$ for the case of two right handed
neutrinos. Measurement of $\theta_{13}$ will have important implications
for both the models; in particular we show that in a class of models with
two right handed neutrinos with approximate $\mu-\tau$ symmetry breaking,
there is a lower limit on $\theta_{13}$, which is between 0.1 to 0.15 
depending on the values of the CP phase. These values are in the range
which will be probed in experiments in near future\cite{bnl}.

The basic assumption under which the two results are derived are the
following:

(A) type I seesaw formula is responsible for neutrino masses:

(B) $\mu-\tau$ symmetry for leptons is broken only at high scale in the
mass matrix of the right handed neutrinos.

The paper is organized as follows: in sec. II, we outline the
general framework for our discussion; in sec. III, we rederive the result 
of ref.\cite{nasri} for the case of exact $\mu-\tau$ symmetry; in sec.
IV, we derive the connection between $\epsilon_l$ and oscillation
parameters for the case of approximate $\mu-\tau$ symmetry. Sec. IV is 
devoted to the case of two right handed neutrinos, where we present the 
allowed range of $\theta_{13}$ dictated by leptogenesis argument.
In sec. V, we describe
a class of simple gauge models where these conditions are
satisfied.

\section{Introductory remarks on lepton asymmetry in type I seesaw models}
We start with an extension of the minimal supersymmetric standard
model (MSSM) for the generic the type I  seesaw model for neutrino
masses. The effective low energy superpotential for this model is
given by
\begin{eqnarray}
W~=~e^{cT}{\bf Y_{\ell}}L H_d+ {N^c}^T{\bf Y_{\nu}}L H_u+ {\bf
\frac{M_R}{2}{N^c}^TN^c}
\end{eqnarray}
Here $ L, e^c, \nu^c$ are leptonic superfields; $H_{u,d}$ are the
 Higgs fields of MSSM. $Y_\nu$ and $M_R$ are general matrices where we
choose a basis where $Y_\ell$ is diagonal. We do not display the
quark part of the superpotential which is same as in the MSSM.
After electroweak symmetry breaking, this leads to the type I
seesaw formula for neutrino masses given by
 \begin{eqnarray}
{\cal M}_{\nu}~=~-{\bf Y^T_\nu f^{-1}Y_{\nu}}
\frac{v^2_{wk}tan^2\beta}{v_R} \label{seesaw}\end{eqnarray} The
constraints of $\mu-\tau$ symmetry will manifest themselves in the
form of the $Y_\nu$ and $M_R$. It has been pointed out that if we
go to a basis where the right handed neutrino mass matrix is
diagonal, we can solve for $Y_\nu$ in terms of the neutrino
masses and mixing angles as follows\cite{casas}:
\begin{eqnarray}
Y_\nu v~=~i{M^d_R}^{1/2}R(z_{ij})({\cal
M}^d_\nu)^{1/2}U^{\dagger} \label{ynu}\end{eqnarray} where $R$ is
a complex matrix with the property that $RR^T=1$. The unitary
matrix $U$ is the lepton mixing matrix defined by
\begin{eqnarray}
{\cal M}_\nu~=~U^*{\cal M}^d_\nu U^{\dagger}
\end{eqnarray}
The complex orthogonal matrices $R$ can be parameterized as:
\begin{eqnarray}
R(z_{12}, z_{23}, z_{13})~=~R(z_{23})R(z_{13})R(z_{12})
\end{eqnarray}
with
\begin{eqnarray}
R(z_{12})~=~\pmatrix{cos z_{12} & sin z_{12} & 0\cr -sin
z_{12} &
cos z_{12} & 0 \cr 0 & 0 & 1}
\end{eqnarray}
 and similarly for the other matrices.
$z_{ij}$ are complex angles.

Let us now turn to lepton asymmetry:
the formula for primordial lepton asymmetry in this case, caused by
right handed
neutrino decay is
\begin{eqnarray}
\epsilon_l~=~\frac{1}{8\pi}\sum_j\frac{ Im [\tilde{Y}_\nu
\tilde{Y}^{\dagger}_\nu]^2_{1j}}{(\tilde{Y}_\nu
\tilde{Y}^{\dagger}_\nu)_{11}}
F(\frac{M_1}{M_j})
\label{nl}
\end{eqnarray}
where $\tilde{Y}_\nu$ is defined in a basis where righthanded neutrinos
are mass eigenstates and their masses are denoted by $M_{1,2,3}$ where
$F(x)~=~-\frac{1}{x}\left[\frac{2x^2}{x^2-1}-ln({1+x^2})\right]$\cite{vissani}.
In the case where that the right
handed neutrinos have a hierarchical mass pattern i.e. $M_1 \ll
M_{2,3}$, we get  $F(x)~\simeq -{3} x$. In this
approximation, we can write the lepton asymmetry in a simple
form\cite{buch2}
\begin{eqnarray}
\epsilon_l ~=~-\frac{3}{8\pi}\frac{M_1 Im [Y_\nu {\cal M}^{\dagger}_\nu
Y^T_\nu]_{11}}{v^2(\tilde{Y}_\nu
\tilde{Y}^{\dagger}_\nu)_{11}}
\label{buch}
\end{eqnarray}
where Using the expression for $Y_\nu$ given above, we can rewrite
$\epsilon_l$
as:
\begin{eqnarray}
\epsilon_l~=~-\frac{3}{8\pi}\frac{ Im [{M^d_R}^{1/2}R(z_{ij})
{{\cal M}^d}^2_\nu R(z_{ij}) {M^d_R}^{1/2}
]_{11}}{v^2|R(z_{ij}){\cal M}_\nu
R^{\dagger}(z_{ij})|^2_{11}} \label{buch1}\end{eqnarray}

We will now apply this discussion to calculate the lepton
asymmetry in the general case without any symmetries. In the
following sections, we follow it up with a discussion of two
cases: (i) the cases of exact $\mu-\tau$ symmetry and (ii) the
case where this symmetry is only approximate. Since the formula
in Eq. (9) assumes that there are three right handed neutrinos, we
will focus on this case in the next two sections. In a subsequent
section, we consider the case of two right handed neutrinos
$(N_\mu,N_\tau)$, which transform into each other under the
$\mu-\tau$ symmetry. Both cases are in agreement with the
observed neutrino mass differences and mixings.

It follows from Eq.\ref{buch1} that
\begin{eqnarray}
\epsilon_l~=~-\frac{3M_1}{8\pi}\frac{Im[m^2_1 R^2_{11}+m^2_2
R^2_{12}+m^2_3 R^2_{13}]}{v^2|R(z_{ij}){\cal M}_\nu
R^{\dagger}(z_{ij})|^2_{11}} \label{for}
\end{eqnarray}
 Since the
matrix $R$ is an orthogonal matrix, we have the relation
\begin{eqnarray}
R^2_{11}+R^2_{12}+R^2_{13}~=~1
\end{eqnarray}
Using this equation in Eq.\ref{for}, we get
\begin{eqnarray}
\epsilon_l~=~-\frac{3M_1}{8\pi}\frac{Im[\Delta m^2_\odot
R^2_{12}+\Delta m^2_A R^2_{13}]}{v^2\sum_j(|R_{1j}|^2 m_j)}
\label{for1}
\end{eqnarray}
This relation connects the lepton asymmetry to both the solar and
the atmospheric mass difference square\cite{BPD}. To make a
prediction for the lepton asymmetry, we need to the lengths of
the complex quantities $R_{1j}$. The out of equilibrium condition
does provide a constraint on $|R_{1j}|$ as follows:
\begin{eqnarray}
\sum_{j=1,2,3}(|R_{1j}|^2 m_j)\leq 10^{-3} ~eV
\label{equ1}
\end{eqnarray}
It is clear from Eq. (13) that if neutrinos are quasidegenerate
i.e. $m_1\simeq m_2 \simeq m_3 \equiv m_0$, then using Eq. (11),
we find that the left hand side of Eq. (13) has a lower bound of
$m_0$ which is clearly much bigger than the right hand side of the 
inequality. Defining $K \equiv \frac{\Gamma}{H}$, this means that $K\geq 
\frac{m_0}{2\times 10^{-3}~ eV}\gg 1$. This implies that the right handed 
neutrinos decays are in equilibrium at $T\simeq M_1$. This will cause 
dilution of the lepton asymmetry generated with the dilution factor given 
by $K$. Using a parameterization for the dilution factor
$\kappa_1\simeq \frac{0.3}{K(ln K)^{3/5}}$\cite{dilution}, we get 
$\kappa_1\simeq
10^{-3}$ which will make the baryon to photon ratio much too
small. Based on this argument, we conclude that a degenerate mass
spectrum with $m_0\geq 0.1$ eV will most likely be in conflict
with observations, if type I seesaw is responsible for neutrino masses. 
It must however be noted that a more appealing and natural scenario for
degenerate neutrino masses is type II seesaw formula\cite{seesaw2}, in 
which case the above considerations do not apply. Therefore, it is not 
possible to conclude based on the leptogenesis argument alone that a 
quasi-degenerate neutrino spectrum is inconsistent.

In a hierarchical neutrino mass
picture, Eq. (13) implies that $|R_{13}|^2 \leq 0.02$ and $|R_{12}|^2
\leq 0.1$. If we assume that the upper limit in the Eq.\ref{equ1}
is saturated, then we get the atmospheric neutrino mass
difference square in Eq.\ref{for1} to give the dominant
contribution. We will see below that if one assumes an exact
$\mu-\tau$ symmetry for the neutrino mass matrix, the situation
becomes different and it is the solar mass difference square that
dominates.

\section{Three right handed neutrinos and exact $\mu-\tau$ symmetry}
In this section, we consider the case of three right handed
neutrino with an exact $\mu-\tau$ symmetry in the Dirac mass
matrix as well as the right handed neutrino mass matrix. In this
case, the right handed neutrino mass matrix $M_R$ and the Dirac
Yukawa coupling $Y_\nu$ can be written respectively as:
\begin{eqnarray}
{\bf M_R}~=~\left(\begin{array}{ccc}M_{11} & M_{12} & M_{12}\\ M_{12} &
M_{22} &
M_{23}\\ M_{12} & M_{23} & M_{22}\end{array}\right)\\ \nonumber
{\bf Y_\nu}~=~\left(\begin{array}{ccc}h_{11} & h_{12} & h_{12}\\ h_{21} &
h_{22} & h_{23} \\
h_{21} & h_{23} & h_{22}\end{array}\right)
\end{eqnarray}
where $M_{ij}$ and $h_{ij}$ are all complex. An important property of
these two matrices
is that they can be cast into a block diagonal form by the same
transformation
matrix $U_{23}(\pi/4)\equiv \left(\begin{array}{cc} 1 & 0\\ 0 &
U(\pi/4)\end{array}\right)$ on the $\nu$'s and $N$'s. Let us denote the
block diagonal forms by a
tilde i.e. $\tilde{Y}_\nu$ and $\tilde{M}_R$. We then go to a basis where
the $\tilde{M}_R$ is subsequently
diagonalized by the most general $2\times 2$ unitary matrix as
follows:
\begin{eqnarray}
V^T(2\times 2)U^T_{23}(\pi/4)M_RU_{23}(\pi/4)V(2\times 2)~=~ M^d_R
\end{eqnarray}
where $V(2\times 2)~=~\left(\begin{array}{cc}V & 0\\ 0 &
1\end{array}\right)$ where $V$ is the most general $2\times 2$
unitary matrix given by $V= e^{i\alpha}P(\beta)R(\theta)P(\gamma)$.
The $3\times 3$ case therefore reduces to a $2\times 2$ problem. The third
mass eigenstate in both the light and the heavy sectors play no role in
the leptogenesis as well as generation of solar mixing angle\cite{nasri}.
Note also that we have $\theta_{13}=0$. The seesaw formula in the 1-2
subsector has exactly the same form except that all matrices in the
left and right hand side of Eq. (9) are $2\times 2$ matrices.
The formula for the Dirac Yukawa coupling in this case can be inverted
to the form:
\begin{eqnarray}
\tilde{Y}_\nu(2\times 2)~=~i{M^d_R}^{1/2}(2\times
2)R(z_{12})({\cal M}^d_\nu)^{1/2}(2\times 2)\tilde{U}^{\dagger}
\label{ynu1}\end{eqnarray} where
$U~=~U_{23}(\pi/4)\pmatrix{\tilde{U} & 0 \cr 0 & 1}$. Using this,
we can cast $\epsilon_l$ in the form:
\begin{eqnarray}
\epsilon_l~=~\frac{3}{8\pi}\frac{M_1}{v^2}\frac{Im(cos^2~z_{12})
\Delta m^2_\odot}{(|cos z_{12}|^2 m_1+|sin z_{12}|^2 m_2)}
\end{eqnarray}
This could also have been seen from Eq.(\ref{for1}) by realizing
that for the case of exact $\mu-\tau$ symmetry, we have $z_{13}=0$
and $z_{23}=\pi/4$.

 The above result reproduces the direct
proportionality between $\epsilon_l$ and solar mass difference
square found in Ref.\cite{nasri}. To simplify this expression
further, let us note that out of equilibrium condition for the
decay of the lightest right handed neutrino leads to the
condition:
\begin{eqnarray}
\frac{M^2_1}{v^2_{wk}}[m_1|cos z_{12}|^2+ m_2 |sin z_{12}|^2]
\leq 14 \frac{M^2_1}{M_{P\ell}}
\end{eqnarray}
which implies that
\begin{eqnarray}
|m_1|cos z_{12}|^2+ m_2 |sin z_{12}|^2| \leq 2\times 10^{-3}~eV
\label{equ}\end{eqnarray}
 Since solar neutrino data require that
in a hierarchical neutrino mass picture $m_2 \simeq 0.9\times
10^{-2}$ eV, in Eq.(\ref{equ}), we must have $|sin z_{12}|^2\sim
0.2$. If we parameterize $cos^2~z_{12}~=~ \rho e^{i\eta}$, we
recover the conclusions of Ref.\cite{nasri}. This provides a
different way to arrive at the conclusions of Ref.\cite{nasri}.

\section{Lepton asymmetry and $\mu-\tau$ symmetry breaking}
In this section, we consider the effect of breaking of $\mu-\tau$
symmetry on lepton asymmetry. Within the seesaw framework, this
breaking can arise either from the Dirac mass matrix for the
neutrinos or from the right handed neutrino sector or both. We
focus on the case, when the symmetry is broken in the right handed
sector only. Such a situation is easy to realize in seesaw models
where the theory obeys exact $\mu-\tau$ symmetry at high scale
(above the seesaw scale) prior to B-L symmetry breaking as we
show in a subsequent section. We will also show that in this case
there is a simple generalization of the lepton asymmetry formula
that we derived in the exact $\mu-\tau$ symmetric case
\cite{nasri}\footnote{Leptogenesis in a
specific $\mu-\tau$ symmetric model where the Dirac Yukawa
coupling has the form $Y_\nu~=~diag(a,b,b)$ has been discussed in
Ref.\cite{gl}. Our discussion applies more generally.}.

In this case the neutrino Yukawa matrix is given in the mass
eigenstates basis of the right handed neutrinos by
\begin{equation}
\tilde{Y_{\nu}} = V^{+}_{1/3}V^{+}_{1/2}V^{+}_{2/3}Y_{\nu}
\end{equation}
where $Y_{\nu}$ is the neutrino Dirac matrix in the flavor basis;
The notation $V^{+}_{i/j}$ denotes a unitary $2\times 2$ matrix
in the $(i,j)$ subspace. In the above equation, $V_{2/3} =
V_{2/3}(\pi/4)$. Now if we substitute for $\tilde{Y_{\nu}}$ the
expression in Eq. \ref{ynu} and use maximal mixing for the
atmospheric neutrino we obtain
\begin{equation}
 \left[ \begin{array}{c c}
\tilde{Y}_{2\times 2}&0\\
0&\tilde{y_3}\\
\end{array} \right] =  V_{1/3}M^{1/2}_R 
R_{1/2}R_{1/3}m^{1/2}_{\nu}U_{1/2}^{+}U_{1/3}^{+}
\end{equation}

Since the $\mu-\tau$ symmetry breaking is assumed to be small and
from reactor neutrino experiments $\theta_{13} <<1$ we will expand
the mixing matrices in the $1-3$ subspace to first order in mixing
parameter:
\begin{equation}
(V, R, U)_{1/3} \simeq 1 + (\epsilon, z, \theta)_{13}E
\end{equation}
where
\begin{equation}
E = \left[ \begin{array}{c c c}
0&0&1\\
0&0&0\\
-1&0&0
\end{array} \right]
\end{equation}
To first order in $\epsilon_{13}$, $z_{13}$ and $\theta_{13}$ we
have
\begin{equation}
z_{13}M^{1/2}_RR_{1/2}Em_{\nu}U^{+}_{1/2} +
\epsilon_{13}EM^{1/2}_RR_{1/2}m^{1/2}_{\nu}U^{+}_{1/2}\\
- \theta_{13}M^{1/2}_RR_{1/2}m^{1/2}_{\nu}U^{+}_{1/2}E = 0
\end{equation}
It is straight forward to show that the perturbation parameters
should satisfy the following equations
\begin{eqnarray}
\epsilon_{13}M_{R_3}m_3 + z_{13}M_{R_1}m_3R_{11} -
\theta_{13}e^{-i\delta}M_{R_1}c_{\theta}(m_1R_{11} - m_2R_{12}) &
\simeq & 0 , \nonumber \\
\epsilon_{13}M_{R_2}(m_2R_{12}s_{\theta} - m_1R_{11}c_{\theta}) -
z_{13}M_{R_3}m_1c_{\theta} -  \theta_{13}e^{-i\delta}M_{R_3}m_3 &
\simeq & 0 , \nonumber \\
\epsilon_{13}M_{R_2}(m_1R_{11}s_{\theta} + m_2R_{12}c_{\theta}) +
z_{13}M_{R_3}m_1s_{\theta} & \simeq & 0, \nonumber \\
z_{13}M_{R_2}m_3R_{21} -
\theta_{13}e^{-i\delta}M_{R_2}c_{\theta}(m_1R_{21} - m_2R_{22}) &
\simeq & 0
\end{eqnarray}
Where $R_{ij}$ are the matrix elements of $R_{1/2}$ and
$c_{\theta}$ and $s_{\theta}$ are the sine and cosine of the solar
neutrino mixing angle. Hence one can see that the parameter
$z_{13}$ is proportional to the $\theta_{13}$ neutrino mixing
angle and is given to first order by
\begin{equation}
z_{13} = [(\frac{m_1}{m_3})R_{21} -
(\frac{m_2}{m_3})R_{22}]\theta_{13}e^{-i\delta}c_{\theta}
\end{equation}
This proves that the matrix element $R_{13}$ that goes into the
leptogenesis formula is directly proportional to the physically
observable parameter $\theta_{13}$. This enables us to write 
$\epsilon_l=a\Delta m^2_\odot + b \Delta m^2_A \theta^2_{13}$.  A 
consequence of this is that
if the coefficient of proportionality is chosen to be of order
one, then as experimental upper limit goes down, unlike the
generic type I seesaw case in section II, the solar mass
difference square starts to dominate for the LMA solution to the
solar neutrino problem.

\section{Lepton asymmetry for two right handed neutrinos}
In this section, we consider the case of two right handed
neutrinos which transform into one another under $\mu-\tau$
symmetry. The leptogenesis in this model with exact $\mu-\tau$
symmetry was discussed in \cite{nasri} and was shown that it
vanishes. In this model therefore, a vanishing or very tiny
$\theta_{13}$ would not provide a viable model for leptogenesis.
Turning this argument around, enough leptogenesis should provide
a lower limit on the value of $\theta_{13}$.

To set the stage for our discussion, let us first review the
argument for the exact $\mu-\tau$ symmetry case\cite{nasri}. The
symmetry under which $(N_\mu\leftrightarrow N_\tau)$ and
$L_\mu\leftrightarrow L_\tau$ whereas the $m_\mu\neq m_\tau$
constrains the general structure of ${\bf Y_\nu}$ and ${\bf M_R}$
as follows:
\begin{eqnarray}
{\bf M_R}~=~\left(\begin{array}{cc}M_{22} & M_{23}\\ M_{23} &
M_{22}\end{array}\right)\\ \nonumber
{\bf Y_\nu}~=~\left(\begin{array}{ccc}h_{11} & h_{22} &  h_{23} \\
h_{11} & h_{23} & h_{22}\end{array}\right)
\end{eqnarray}
 In order to calculate the lepton asymmetry using Eq.(7), we first
diagonalize the righthanded neutrino mass matrix and change the
$Y_\nu$ to $\tilde{Y}_\nu$. Since ${\bf M_R}$ is a symmetric
complex $2\times 2$ matrix, it can be diagonalized by a
transformation matrix $U(\pi/4)\equiv\frac{1}{\sqrt{2}}
\left(\begin{array}{cc}1 & 1\\ -1 & 1\end{array}\right)$ i.e.
$U(\pi/4){\bf M_R}U^T(\pi/4)~=~diag(M_1 , M_2)$ where $M_{1,2}$
are complex numbers. In this basis we have $\tilde{\bf
Y}_\nu~=~U(\pi/4){\bf Y_\nu}$. We can therefore rewrite the
formula for $n_\ell$ as
\begin{eqnarray}
\epsilon_l\propto \sum_j Im [U(\pi/4){\bf Y}_\nu {\bf
Y}^{\dagger}_\nu U^T(\pi/4)]^2_{12} F(\frac{M_1}{M_2})
\end{eqnarray}
Now note that ${\bf Y}_\nu {\bf Y}^{\dagger}_\nu$ has the form
$\left(\begin{array}{cc}A & B\\ B & A\end{array}\right)$ which
can be diagonalized by the matrix $U(\pi/4)$. Therefore it
follows that $\epsilon_\ell =0$.

Let us now introduce $\mu-\tau$ symmetry breaking. If we introduce
a small amount of $\mu-\tau$ breaking in the right handed
neutrino sector as follows: we keep the $Y_\nu$ symmetric but choose
the right handed neutrino mass matrix as:
\begin{eqnarray}
{\bf M_R}~=~\left(\begin{array}{cc}M_{22} & M_{23}\\ M_{23} &
M_{22}(1+\beta)\end{array}\right).
\end{eqnarray}
After the right handed neutrino mass matrix is
diagonalized, the $3\times 2$ $Y'_\nu$ takes the form (for
$\theta_{13} \ll 1$ and in the basis where the light neutrino masses are 
diagonal):
\begin{eqnarray}
\left(\matrix{A&B&w\theta_{13}\cr
x\theta_{13}&y\theta_{13}&D}\right)
\end{eqnarray}
Here $B, D, x, y, w$ are of order one and $\theta_{13}\propto \beta$.

To first order in the small mixing $\theta_{13}$, the complex parameters 
$A,B,D$ satisfy the constraint
\begin{eqnarray}
A\sim \theta_{13};~~Bv^2\simeq m_2 M_1; ~~ Dv^2\simeq m_3 M_2
\end{eqnarray}
Using these order of magnitude values, we now find that
\begin{eqnarray}
\epsilon _l \simeq
 \frac{3}{8\pi}\frac{M_1}{v^2}
\frac{\sin\eta[m^2_3\theta^2_{13}\xi]}{m_2}
\end{eqnarray}
where $\xi$ is a function of order one.
It is clear that very small values for $\theta_{13}$ will lead to
 unacceptably small $\epsilon_l$. In Fig. 1, we have plotted
$\eta_B$ against $\theta_{13}$ for values of the parameters
in the model that fit the oscillation data and find a lower bound
on $\theta_{13} \geq 0.1-0.15 $ for two different values of the CP 
phases (figure 1). In this figure, we have chosen, $M_1\simeq 7\times 
10^{11}$ GeV. For higher values of $M_1$ the allowed range $\theta_{13}$ 
moves to the lower range. Also we note that for values of $M_1 < 
7\times 10^{11}$ GeV, the baryon asymmetry becomes lower than the observed 
value.

\begin{figure}[h]
\begin{center}
\includegraphics[scale=1.3]{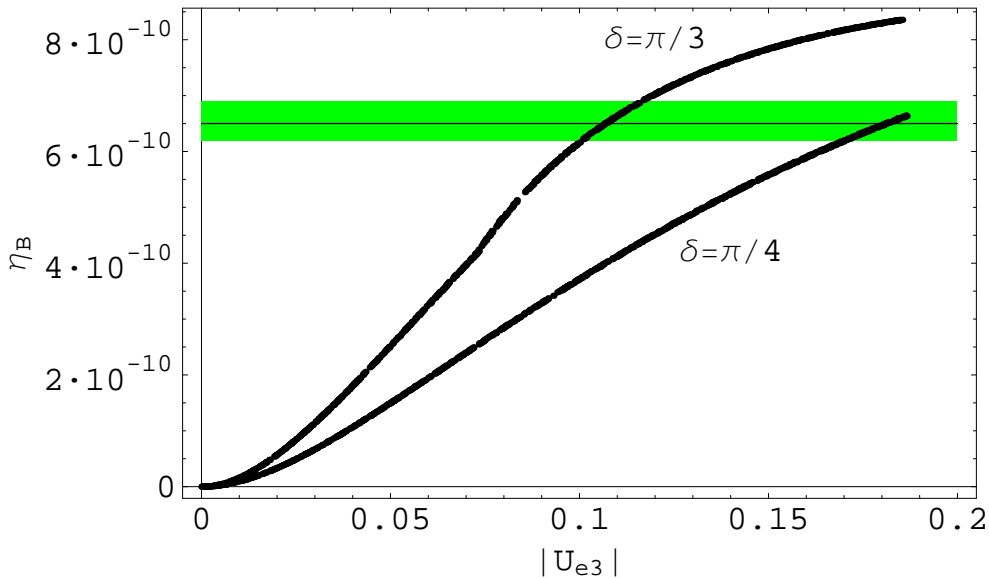}
\caption{Plot of $\eta_B$ vrs $\theta_{13}$ for the case of
two right handed neutrinos with approximate $\mu-\tau$ symmetry and CP 
phases $\delta = \pi/4$ and $\pi/3$. The values of $\theta_{13}$ are 
predicted to be 0.1 and 0.15 respectively. The horizontal line 
corresponds to $\eta^{obs}_B~=~(6.5^{+0.4}_{-0.3})\times 
10^{-10}$\cite{wmap}.}
\end{center}
\end{figure}

\section{A model for $\mu-\tau$ symmetry for neutrinos}

In this section, we present a simple extension of the minimal
supersymmetric standard model (MSSM) by adding to it specific high scale
physics that at low energies can exhibit $\mu-\tau$ symmetry in the
neutrino sector as well as real Dirac masses for neutrinos.

First we recall that MSSM needs to be extended by the addition of
a set of right handed neutrinos (either two or three) to
implement the seesaw mechanism for neutrino masses\cite{seesaw}.
We will accordingly add three right handed neutrinos
$(N_e,N_\mu,N_\tau)$ to MSSM. We then assume that at high scale,
the theory has $\mu-\tau$ $S_2$ symmetry under which
$N_{\pm}\equiv (N_\mu\pm N_\tau)$ are even and odd combinations;
similarly, we have for leptonic doublet superfields
$L_{\pm}\equiv (L_\mu\pm L_\tau)$ and leptonic singlet ones
$\ell^c_{\pm} \equiv (\mu^c\pm \tau^c)$;  two pairs of Higgs
doublets ($\phi_{u,\pm}$ and $\phi_{d,\pm}$), and a singlet
superfields $S_{\pm}$. Other superfields of MSSM such as
$N_e,L_e,e^c$ as well as quarks are even under the $\mu-\tau$
$S_2$ symmetry. Now suppose that we write the superpotential
involving the $S$ fields as follows:
\begin{eqnarray}
W_S~=~\lambda_1 \phi_{u,-}\phi_{d,+}S_-+\lambda_2 \phi_{u,-}\phi_{d_-}S_+
\end{eqnarray}
then when we give high scale vevs to $<S_{\pm}>~=~M_{\pm}$, then below the
high scale there are only the usual MSSM Higgs pair $H_u\equiv~\phi_{u,+}$
and $H_d~\equiv~(c\phi_{d,+}+s \phi_{d,-})$ that survive whereas the
other pair becomes superheavy and decouple from the low energy Lagrangian.
The effective coupling at the MSSM level is then given by:
\begin{eqnarray}
W~=~h_e L_eH_{d}e^c+h_1L_eH_{d}\ell^c_{+} +h_2L_eH_{d}m^c_{-}
+h_3L_{+}H_{d}e^c\\ 
\nonumber
+h_4L_-H_{d}e^c+h_5L_+H_{d}\ell^c_{+}+h_6L_-H_{d}m^c_{-}
+h_7L_-H_{d}\ell^c_{+}\\ \nonumber +f_1L_eH_{u,+}N_{e}+
f_2L_eH_{u,+}N_{+}+f_3L_+H_{u,+}N_{e}+f_4L_+H_{u,+}N_{+}\\ 
\nonumber
+f_5L_-H_{u,+}N_{-}
\end{eqnarray}
Note that the $\mu-\tau$ symmetry is present in the Dirac neutrino mass
matrix whereas it is not in the charged lepton sector as would be
required to .

We show below that it is possible to have a high scale
supersymmetric theory which would lead to real Dirac Yukawa
couplings ($f_{i}$) if we require the high scale theory to be
left-right symmetric. To show how this comes about, consider the
gauge group to be $SU(2)_L\times SU(2)_R \times U(1)_{B-L}$ with
quarks and leptons assigned to left and right handed doublets as
usual\cite{lr} i.e. $Q(2,1,1/3)$, $Q^c(1,2,-1/3)$; $L(2,1, -1)$
and $L^c(1,2,+1)$; Higgs fields $\Phi(2,2,0)$; $\chi (2,1,+1)$;
$\bar{\chi}(2,1,-1)$; $\chi^c(1,2,-1)$ and $\bar{\chi}_c(1,2,-1)$.
The new point specific to our model is that we have two sets of
the Higgs fields with the above quantum numbers, one even and the
other odd under the $\mu-\tau$ $S_2$ permutation symmetry i.e.
$\Phi_\pm $, $\chi_\pm $, $\bar{\chi}_\pm $, $\chi^c_\pm $ and
$\bar{\chi}^c_\pm $ (plus for fields even under $S_2$ and $-$ for
fields odd under $S_2$. ) Furthermore, we will impose the parity
symmetry under which $Q\leftrightarrow {Q^c}^*$,
$L\leftrightarrow {L^c}^*$, $(\chi,\bar{\chi}\leftrightarrow
{\chi^c}^*, {\bar{\chi^c}}^*$), $\Phi\leftrightarrow
{\Phi}^{\dagger}$.

The Yukawa couplings of this theory invariant under the gauge
group as well as parity are given by the superpotential:
\begin{eqnarray}
W~=~h_{11}L^T_e\Phi_+L^c_e~+~h_{++}L^T_+\Phi_+L^c_+~h_{--}L^T_-\Phi_+L^c_-
~h_{e+}L^T_e\Phi_+L^c_+~
+~h^*_{e+}L^T_+\Phi_+L^c_e~\\
\nonumber+~h_{e-}L^T_e\Phi_-L^c_-+~h^*_{e-}L^T_-\Phi_-L^c_e
+~h_{+-}L^T_+\Phi_-L^c_-+~h^*_{+-}L^T_-\Phi_-L^c_\\
\nonumber
\end{eqnarray}
where $h_{11},h_{++},h_{--}$ are real.

The Higgs sector of the low energy superpotential is determined
from this theory after left-right gauge group is broken down to
the standard model gauge group by the vev's of $\chi^c$. The
phenomenon of doublet-doublet spitting leaves only two Higgs
doublets out of the four in $\Phi_{\pm}$ and is determined by a
generic superpotential of type
\begin{eqnarray}
W_{DD}~=~\sum_ {i,j,k}\lambda_{ijk}\chi_i\Phi_j\chi^c_k
+\lambda'_{ijk}\bar{\chi}_i\phi_j\bar{\chi}^c_k~+
M_1(\chi_{\pm}\bar{\chi}_{\pm}+\chi^c_\pm\bar{\chi^c}_\pm)
\end{eqnarray}
where $i,j,k$ go over $+$ and $-$ for even and odd and only even
terms are allowed by $\mu-\tau$ invariance e.g. $\lambda_{+++},
\lambda_{+--},...$ are nonzero. Now suppose that  $<\chi^c_+>=0$
but $<\chi^c_+\->\neq 0$ and $<\bar{\chi^c}_\pm>\neq 0$. These
vevs break the left-right group to the standard model gauge
group. It is then easy to see that below the $<\chi^c>$ scale,
there are only one Higgs pair where $H_u=\phi_{u,+}$ and
$H_d~=~\sum_{i=+,-,3,4}a_i \phi_{d,i}$. Here we have denoted the
$\Phi\equiv (\phi_u,\phi_d)$ and $\phi_{d,3,4}~=~\chi_{\pm}$. The
upshot of all these discussions is that the right handed neutrino
Yukawa couplings are $\mu-\tau$ even and therefore have the form:
\begin{eqnarray}
Y_{\nu}~=~\pmatrix{h_{11} & h_{e+} & 0\cr h^*_{e+} & h_{++} & 0\cr
0 & 0 & h_{--}}
\end{eqnarray}
It is easy to see that redefining the fields appropriately, we
can make $Y_\nu$ real. So the only source of complex phase in
this model is in the RH neutrino mass matrix, which in this model
are generated by higher dimensional couplings of the form
$L^cL^c\bar{\chi^c}\bar{\chi^c}$ as we discuss now.

The most general nonrenormalizable interactions that can give
rise to right handed neutrino masses are of the form:
\begin{eqnarray}
W_{NR}~=~ 
\frac{1}{M}[(L^c_e\bar{\chi^c}_+)^2+L^c_e\bar{\chi^c}_-)^2+(L^c_+
\bar{\chi^c}_+)^2\\ \nonumber
(L^c_-\bar{\chi^c}_-)^2+(L^c_-\bar{\chi^c}_+)^2~+(L^c_+\bar{\chi^c}_-)^2
\\ \nonumber (L^c_+\bar{\chi^c}_-)(L^c_-\bar{\chi^c}_+)
\end{eqnarray}
Note that since both $\bar{\chi^c}_\pm$ acquire vevs, the last
term in the above expression will give rise to $\mu-\tau$
breaking in the RH neutrino sector while preserving it in the
$Y_\nu$. The associated couplings in the above equations are in
general complex. This leads to a realistic three generation model
with approximate $\mu-\tau$ symmetry as analyzed in the previous
sections.

In summary, we have studied the implications for leptogenesis in
models where neutrino masses arise from the type I seesaw
mechanism and where the near maximal atmospheric mixing angle
owes its origin to an approximate $\mu-\tau$ symmetry. We derive
a relation of the form $\epsilon_l = (a \Delta m^2_\odot+ b \Delta
m^2_A \theta^2_{13})$ for the case of three right handed neutrinos, which 
directly connects the neutrino
oscillation parameters with the origin of matter. We also show
that if $\theta_{13}$ is very small or zero, only the LMA
solution to the solar neutrino puzzle would provide an
explanation of the origin of matter within this framework.
Finally for the case of two right handed neutrinos with approximate 
$\mu-\tau$ symmetry, we predict values for $\theta_{13}$ in the range 
$0.1-0.15$ for specific choices of the the high energy phase between 
$\pi/4$ and $\pi/3$.

 This work is supported by the National Science Foundation grant
no. Phy-0354401

\end{document}